\documentclass[twocolumn,pre]{revtex4}

\usepackage{dcolumn}
\usepackage{amsmath}

\usepackage{graphicx}

\begin{document}

\renewcommand{\d}{{\rm d}}
\renewcommand{\O}{{\rm O}}
\newcommand{\e}{{\rm e}}
\newcommand{\half}{\mbox{$\frac12$}}
\newcommand{\set}[1]{\lbrace#1\rbrace}
\newcommand{\av}[1]{\langle#1\rangle}
\newcommand{\eref}[1]{(\ref{#1})}
\newcommand{\etal}{{\it{}et~al.}}
\newcommand{\kin}{k_\mathrm{in}}
\newcommand{\kout}{k_\mathrm{out}}
\newcommand{\jin}{j_\mathrm{in}}
\newcommand{\jout}{j_\mathrm{out}}

\newlength{\figurewidth}
\ifdim\columnwidth<10.5cm
  \setlength{\figurewidth}{0.95\columnwidth}
\else
  \setlength{\figurewidth}{10cm}
\fi
\setlength{\parskip}{0pt}
\setlength{\tabcolsep}{6pt}

\title{Why social networks are different from other types of networks}
\author{M. E. J. Newman}
\author{Juyong Park}
\affiliation{Department of Physics and Center for the Study of Complex
Systems,\\
University of Michigan, Ann Arbor, MI 48109}
\affiliation{Santa Fe Institute, 1399 Hyde Park Road, Santa Fe, NM 87501}
\begin{abstract}
We argue that social networks differ from most other types of networks,
including technological and biological networks, in two important ways.
First, they have non-trivial clustering or network transitivity, and
second, they show positive correlations, also called assortative mixing,
between the degrees of adjacent vertices.  Social networks are often
divided into groups or communities, and it has recently been suggested that
this division could account for the observed clustering.  We demonstrate
that group structure in networks can also account for degree correlations.
We show using a simple model that we should expect assortative mixing in
such networks whenever there is variation in the sizes of the groups and
that the predicted level of assortative mixing compares well with that
observed in real-world networks.
\end{abstract}
\maketitle

\section{Introduction}
\label{intro}
The last few years have seen a burst of interest within the statistical
physics community in the properties of networked systems such as the
Internet, the World Wide Web, and social and biological
networks~\cite{Strogatz01,AB02,DM03b,Newman03d}.  Researchers' attention
has, to a large extent, been focused on properties that seem to be common
to many different kinds of networks, such as the so-called ``small-world
effect'' and skewed degree distributions~\cite{WS98,BA99b,ASBS00}.  In this
paper, by contrast, we highlight some apparent differences between
networks, specifically between social and non-social networks.  Our
observations appear to indicate that social networks are fundamentally
different from other types of networked systems.

We focus on two properties of networks that have received attention
recently.  First, we consider degree correlations in networks.  It has been
observed that the degrees of adjacent vertices in networks are positively
correlated in social networks but negatively correlated in most other
networks~\cite{Newman02f}.  Second, we consider network transitivity or
clustering, the propensity for vertex pairs to be connected if they share a
mutual neighbor~\cite{WS98}.  We argue that the level of clustering seen in
many non-social networks is no greater than one would expect by chance,
given the observed degree distribution.  For social networks however,
clustering appears to be far greater than we expect by chance.

We conjecture that the explanation for both of these phenomena is in fact
the same.  Using a simple network model, we argue that if social networks
are divided into groups or communities, this division alone can produce
both degree correlations and clustering.

The outline of the paper is as follows.  In Sec.~\ref{degcorr} we discuss
the phenomenon of degree correlation and summarize some empirical results
for various networks.  In Sec.~\ref{clustering} we do the same for
clustering.  We also present theoretical arguments that suggest that the
clustering seen in non-social networks is of about the magnitude one would
expect for a random graph model with parameters similar to real networks.
Then in Sec.~\ref{socnet} we present analytic results for a simple model of
a social network divided into groups.  This model, which was introduced
previously~\cite{Newman03e}, is known to generate high levels of
clustering.  Here we show that it can also explain the presence of
correlations between the degrees of adjacent vertices.  In
Sec.~\ref{examples} we compare the model's predictions concerning degree
correlations against two real-world social networks, of collaborations
between scientists and between businesspeople.  In the former case we find
that the model is in good agreement with empirical observation.  In the
latter we find that it can predict some but not all of the observed degree
correlation, and we conjecture that the remainder is due to true
sociological or psychological effects, as distinct from the purely
topological effects contained in the model.  In Sec.~\ref{concs} we give
our conclusions.

\section{Degree correlations}
\label{degcorr}
In studies of the network structure of the Internet at the level of
autonomous systems, Pastor-Satorras~\etal~\cite{PVV01} have recently
demonstrated that the degrees of adjacent vertices in this network appear
to be anticorrelated.  They measured the mean degree~$\av{k_{\rm nn}}$ of
the nearest neighbors of a vertex as a function of the degree~$k$ of that
vertex, and found that the resulting curve falls off with $k$ approximately
as $\av{k_{\rm nn}} \sim k^{-1/2}$.  Thus, vertices of high degree~$k$ tend
to be connected, on average, to others of low degree, and \textit{vice
versa}.  A simple way of quantifying this effect is to measure a
correlation coefficient of the degrees of adjacent vertices in a network,
defined as follows.

Suppose that $p_k$ is the degree distribution of our network, i.e.,~the
fraction of vertices in the network with degree~$k$, or equivalently the
probability that a vertex chosen uniformly at random from the network will
have degree~$k$.  The vertex at the end of a randomly chosen edge in the
network will have degree distributed in proportion to $kp_k$, the extra
factor of $k$ arising because $k$ times as many edges end at a vertex of
degree~$k$ than at a vertex of degree one~\cite{Feld91,MR95,NSW01}.
Commonly we are interested not in the total degree of the vertex at the end
of an edge, but in the ``excess degree,'' which is the number of edges
attached to the vertex other than the one we arrived along, which is
obviously one less than the total degree.  The properly normalized
distribution of the excess degree is
\begin{equation}
q_k = {(k+1) p_{k+1}\over\sum_k k p_k}.
\label{defsqk}
\end{equation}
We then define the quantity $e_{jk}$, which is the joint probability that a
randomly chosen edge joins vertices with excess degrees $j$~and~$k$.

Now consider a network in which the vertices have given degrees (the value
of the degrees being called the ``degree sequence''), but which is in all
other respects random.  That is, the network is drawn uniformly at random
from the ensemble of all possible networks with the given degree sequence.
This is the so-called configuration
model~\cite{Bollobas80,Luczak92,MR95,NSW01}, which we can use as a handy
null model for testing our results.  In the configuration model the
expected value of the quantity~$e_{jk}$ is simply $e_{jk}=q_jq_k$, and by
its deviation from this value we can quantify the level of degree
correlation present, relative to the null model.  We
define~\cite{Newman02f}
\begin{equation}
r = {1\over\sigma_q^2} \sum_{jk} jk(e_{jk}-q_jq_k),
\label{defsr}
\end{equation}
where $\sigma_q^2 = \sum_k k^2 q_k - \bigl[ \sum_k k q_k \bigr]^2$ is the
variance of the distribution~$q_k$.  The quantity~$r$ will be positive or
negative for networks with positive or negative degree correlations
respectively.  In the ecology and epidemiology literatures these two cases
are called ``assortative'' and ``disassortative'' mixing by degree, and
this nomenclature has been adopted by many physicists also.

The findings of Pastor-Satorras~\etal~\cite{PVV01} discussed above suggest
that the Internet should have a negative value for~$r$, and this indeed is
the case.  The most recent structural measurements of the autonomous-system
graph of the Internet~\cite{Chen02} yield a value of $r=-0.193\pm0.002$.
It now appears that similar results apply to essentially all other networks
\emph{except} social networks.  In Refs.~\cite{Newman02f}
and~\cite{Newman03c} we found that almost all networks seem to be
disassortatively mixed, i.e.,~have negative values of the assortativity
coefficient~$r$, except for social networks, which are normally
assortative.  A small number of networks yield inconclusive results because
the errors on $r$ are bigger than its value, but other than these few, the
pattern appears essentially perfect.

Here we propose that this striking pattern arises because disassortativity
is the natural state for all networks, in a sense that we will make clear
shortly.  Left to their own devices, we conjecture, networks normally have
negative values of~$r$.  In order to show a positive value of~$r$, a
network must have some specific additional structure that favors
assortative mixing.  We suggest in Sec.~\ref{socnet} that division into
communities or groups provides such a structure in social networks.

Our conjecture that most networks will be disassortative is motivated by
the work of Maslov~\etal~\cite{MSZ03}.  Using computer simulations, they
showed that on small networks, disassortative mixing is produced if one
restricts the network topology to having at most one edge between any pair
of vertices.  The same result can be demonstrated analytically as
well~\cite{PN03}.  How small a network need be to show this effect depends
on the degree distribution; to see significant disassortativity, the
highest-degree vertices in the network need to have degree on the order of
$\sqrt{n}$, where $n$ is the total number of vertices, so that there is a
substantial probability of some vertex pairs sharing two or more edges.
(Obviously if there is negligible probability of a double edge occurring
anywhere in the network, then the restriction of having no double edges
will have no effect.)  The Internet is a particularly good example of the
effect, since it has a degree distribution that appears approximately to
follow a power law, $p_k\sim k^{-\alpha}$ with $\alpha$
constant~\cite{FFF99,Chen02}, and the fat tail of the power law produces
many vertices of sufficiently high degree.  However, a number of other
networks also fit the bill: the World Wide Web, peer-to-peer networks, food
webs, neural networks, and metabolic networks all have vertices of
sufficiently high degree, at least in some cases.  In their most common
representations these networks also have only single edges between
vertices, and hence we would expect them to have $r<0$, and calculations of
$r$ from structural data confirm that this is the case~\cite{Newman03c}.

In fact, most networks have only single edges between their vertices.
Although it is possible to have double edges in some networks, in practice
these are usually ignored even where they exist and all edges are
represented as single.  For instance, in the World Wide Web it is possible,
and even common, for a Web page to link twice or more to the same other
page, creating a multiple link.  Such links are however normally recorded
as single by Web crawler programs, and hence any information about multiple
links is lost.  Thus many networks may have single edges only because that
is the way researchers have chosen to represent them, and observed
properties such as disassortativity may be purely a product of this choice
of representation rather than a fundamental law of nature.  Other networks
may truly have single edges---metabolic networks and food-webs are possible
examples of this.

Social networks also usually have only single edges between vertex pairs.
Two people are either acquainted with one another or not---we do not
normally have a concept of being ``doubly acquainted'' with a person.
Nonetheless, the assortativity coefficient~$r$ is positive, and sometimes
very positive, for almost all social networks
measured~\cite{Newman02f,Newman03c}.  This appears to indicate some special
structure in social networks that distinguishes them from other types of
networks.  A revealing clue about what this special structure might be
comes from network transitivity, as we now describe.

\section{Clustering}
\label{clustering}
Watts and Strogatz~\cite{WS98} have pointed out that most networks appear
to have high transitivity, also called clustering.  That is, the presence
of a connection between vertices A and~B, and another between B and~C,
makes it likely that there will also be a connection between A and~C.  To
put it another way, if B has two network neighbors, A~and~C, they are
likely to be connected to one another, by virtue of their common connection
with~B.  In topological terms, there is a high density of triangles, ABC,
in the network, and clustering can be quantified by measuring this density:
\begin{equation}
C = {\mbox{$3\times$ number of triangles on the graph}\over
     \mbox{number of connected triples of vertices}},
\label{defsc}
\end{equation}
where a ``connected triple'' means a vertex connected directly to an
unordered pair of others.  In physical terms, $C$~is the probability,
averaged over the network, that two of your friends will be friends also of
one another.  (This is in fact only one definition of the clustering
coefficient.  An alternative definition, given in~\cite{WS98}, has also
been widely used.  The latter however is difficult to evaluate
analytically, and so we avoid it here.)

The value of the clustering coefficient in the null configuration model can
be calculated in a straightforward fashion~\cite{Newman03b,EMB02}.  Suppose
that two neighbors of the same vertex have excess degrees $j$ and~$k$.  The
probability that one particular edge in the network falls between these two
vertices is $2\times j/(2m)\times k/(2m)$, where $m$ is the total number of
edges in the network.  The total number of edges between the two vertices
in question is $m$ times this quantity, or $jk/(2m)$.  Both $j$ and~$k$ are
distributed according to~\eref{defsqk}, since both vertices are neighbors
of~A and, averaging over this distribution, we then get an expression for
the clustering coefficient:
\begin{equation}
C = {1\over2m}\,\Bigl[\sum_k k q_k\Bigr]^2
  = {1\over n}\,{\bigl[\av{k^2}-\av{k}\bigr]^2\over\av{k}^3},
\label{crg}
\end{equation}
where averages are over all vertices and we have made use of $2m=n\av{k}$.

Normally this quantity goes as $n^{-1}$ and so is very small for large
graphs.  However, some graphs are not large, and hence $C$ is not
negligible.  Consider for example the foodweb of organism in Little Rock
Lake, WI, which was originally analyzed by Martinez~\cite{Martinez91} and
has been widely studied in the networks literature.  This network has
$n=92$, $\av{k}=21.0$, and $\av{k^2}=655.2$.  Plugging these figures into
Eq.~\eref{crg} gives $C=0.47$.  The measured value of $C$ is~$0.40$.  Thus
it appears that we need invoke no special clustering process to explain the
clustering in this network.  Similar results can be found for other small
networks.

This argument can also be applied to some larger networks as well,
particularly those with power-law degree distributions.  The fat tail of
the degree distribution in power-law networks can affect the value of the
clustering coefficient strongly.  To see this consider first how the degree
of the highest-degree vertex in the configuration model varies with system
size~\cite{Newman03d}.

The probability of there being exactly $m$ vertices of degree~$k$ in the
network and no vertices of degree greater than~$k$ is ${n\choose m} p_k^m
(1-P_k)^{n-m}$, where
\begin{equation}
P_k = \sum_{k'=k}^\infty p_{k'},
\label{cumulative}
\end{equation}
is the probability that a vertex has degree greater than or equal to~$k$.
Then the probability~$h_k$ that the highest degree in the network is~$k$ is
\begin{eqnarray}
h_k &=& \sum_{m=1}^n {n\choose m} p_k^m (1-P_k)^{n-m}\nonumber\\
    &=& (p_k+1-P_k)^n - (1-P_k)^n,
\end{eqnarray}
and the expected value of the highest degree is $k_\mathrm{max} = \sum_k k
h_k$.

The value of~$h_k$ tends to zero for both small and large values of~$k$,
and the sum over $k$ is dominated by the terms close to the maximum.  Thus,
in most cases, a good approximation to the expected value of the maximum
degree is given by the modal value.  Differentiating and observing that $\d
P_k/\d k=p_k$, we find that the maximum of~$h_k$ occurs when
\begin{equation}
\biggl( {\d p_k\over\d k} - p_k \biggr) (p_k+1-P_k)^{n-1}
  + p_k (1-P_k)^{n-1} = 0,
\end{equation}
or $k_\mathrm{max}$ is a solution of
\begin{equation}
{\d p_k\over\d k} \simeq -np_k^2,
\end{equation}
where we have made the assumption that $p_k$ is sufficiently small for
$k\gtrsim k_\mathrm{max}$ that $np_k\ll1$ and $P_k\ll1$.  For a degree
distribution with a power-law tail $p_k\sim k^{-\alpha}$, we then find that
\begin{equation}
k_\mathrm{max} \sim n^{1/(\alpha-1)}.
\label{kmax}
\end{equation}
(As shown by Cohen~\etal~\cite{CEBH00}, a simple rule of thumb that leads to
the same result is that the maximum degree is roughly the value of $k$ that
solves $nP_k=1$.)

Most networks of interest have $\alpha<3$, which means $\av{k^2}\sim
k_\mathrm{max}^{3-\alpha}\sim n^{(3-\alpha)/(\alpha-1)}$ and $\av{k}$ is
independent of~$n$.  Then~\eref{crg} gives
\begin{equation}
C \sim n^{(7-3\alpha)/(\alpha-1)}.
\label{cpl}
\end{equation}
If $\alpha>\frac73$, this means that $C$~tends to zero as the graph becomes
large, although it does so slower than the explicit $C\sim n^{-1}$ of
Eq.~\eref{crg}.  At $\alpha=\frac73$, $C$~becomes constant (or logarithmic)
in the graph size.  And remarkably, for $\alpha<\frac73$ it actually
increases with increasing system size, becoming arbitrarily large as
$n\to\infty$.  Thus for $\alpha\lesssim\frac73$, we might expect to see
quite large values of $C$ even in large networks.

Taking the case of the World Wide Web, for example, we find the predicted
value of the clustering coefficient for the configuration model is
$C=0.048$~\cite{Newman03b}, while the measured value is $0.11$---certainly
not perfect agreement, but of the right order of magnitude.  Other examples
err in the opposite direction.  Maslov~\etal~\cite{MSZ03} for instance cite
the example of the Internet, for which they show using numerical
simulations that the observed clustering is actually lower than that
expected for an equivalent random graph model.

It is worth noting that Eq.~\eref{cpl} implies the clustering coefficient
can be greater than~1 if $\alpha<\frac73$.  Physically this means that
there will be more than one edge on average between two vertices that share
a common neighbor.  This is perhaps at odds with the conventional
interpretation of the clustering coefficient as the probability that there
exists \emph{any} edge between the given two vertices---normally one would
not distinguish between the case where there are two edges and the case
where there is one.  (Indeed, as mentioned in Sec.~\ref{degcorr}, in many
networks, one ignores double edges altogether.)  If one takes this
approach, then the value of the clustering coefficient is modified for
networks that would otherwise have $C>1$ as follows.

Consider again two vertices that are neighbors of vertex~A, with excess
degrees $j$ and~$k$.  The probability that a particular edge falls between
them is $2\times j/(2m)\times k/(2m)$, as before, and the probability that
it does not is 1 minus this quantity.  Then the probability that no edge
falls between this pair is
\begin{equation}
\biggl[ 1 - {jk\over2m^2} \biggr]^m \simeq \e^{-jk/2m},
\end{equation}
where the equality becomes exact in the limit of large~$m$.  Thus the
probability of any edge falling between the two vertices is
$1-\e^{-jk/2m}$, and the correct expression for the clustering coefficient
is the average of this:
\begin{equation}
C = \sum_{jk} q_j q_k \bigl( 1-\e^{-jk/2m} \bigr).
\end{equation}
In fact, however, using this expression makes only the smallest of
differences to the expected value of~$C$ on, for example, the World Wide
Web.

All of this demonstrates that for many non-social networks, including
foodwebs, the Internet, and the World Wide Web, clustering can be explained
by a simple random model.  The same however is not true of social networks.
It turns out that social networks in general have a far higher degree of
clustering than the corresponding random model.  We give four examples: the
widely studied network of film-actor collaborations~\cite{WS98,ASBS00},
collaboration networks of mathematicians~\cite{GI95,BM00} and company
directors~\cite{DYB01}, and an email network~\cite{NFB02}.  For these four
networks the theory presented above predicts values of the clustering
coefficient of $0.0098$, $0.00015$, $0.0035$, and $0.017$.  The actual
measured values are $0.20$, $0.15$, $0.59$, and $0.17$, in each case at
least an order of magnitude greater than the prediction.  The implication
appears to be that there is some mechanism producing clustering in social
networks that is not present at a significant level in non-social networks
(or not at least in the examples studied here).  Recent
work~\cite{GN02,Guimera03,RB03,Newman03e,TWH03} suggests a possible
candidate theory, that social networks contain groups, or ``community
structure''~\footnote{An alternative theory is that individuals introduce
pairs of their acquaintances to one another, thus completing network
triangles and increasing the clustering coefficient.  Several models of
this ``triadic closure'' process have been studied in the
literature~\cite{BC96,Watts99b,JGN01,DEB02,KE02,JJ02}.}.

\section{Community structure in networks}
\label{socnet}
In~\cite{Newman03e} one of us proposed a simple model of a network with
community structure and showed that this structure produces substantial
clustering, with values of $C$ that do not go to zero as the network size
becomes large.  Thus the results of the preceding section could be
explained if social networks possess community structure and other types of
networks do not (or they possess it to a lesser degree).  We now show that
the same distinction can also explain the observed difference in degree
correlations between social and non-social networks.

\begin{figure}
\begin{center}
\resizebox{\figurewidth}{!}{\includegraphics{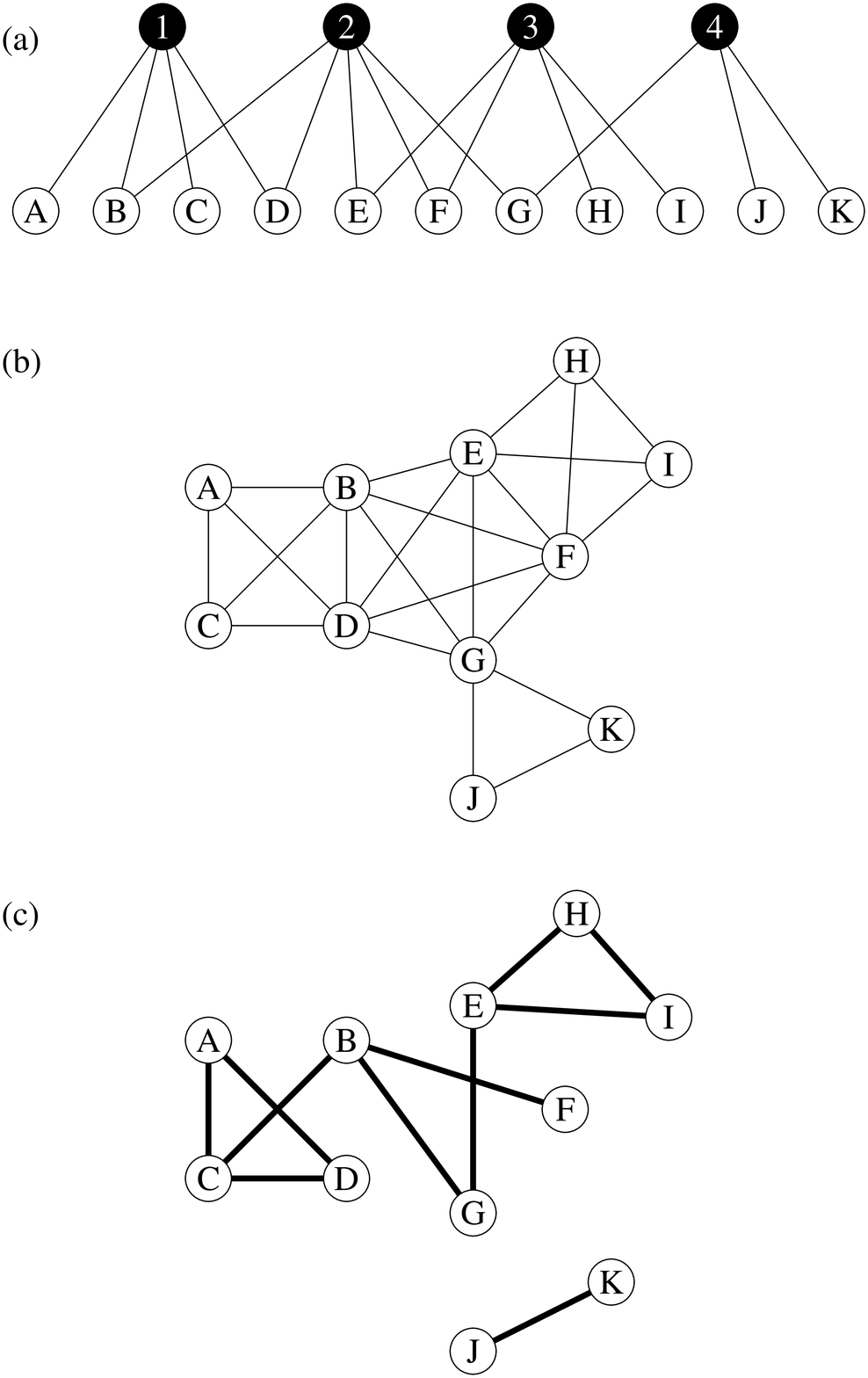}}
\end{center}
\caption{The structure of the network model studied in Sec.~\ref{socnet}.
(a)~We represent individuals (A--K) and the groups (1--4) to which they
belong with a bipartite graph structure.  (b)~The bipartite graph is
projected onto the individuals only, giving a network with edges between
any pair of individuals who share a group.  (c)~The actual social
connections between individuals are chosen by bond percolation on this
projection with bond occupation probability~$p$.  The net result is that
individuals have probability $p$ of knowing others with whom they share a
group.}
\label{model}
\end{figure}

In our model the network is divided into groups and each individual can
belong to any number of groups.  Individuals do not necessarily know all
those with whom they share a group, but instead have probability $p$ of
acquaintance.  They have probability zero of knowing those with whom they
do not share a group.  Mathematically the model can be represented as a
bond percolation process with occupation probability~$p$ on the network
formed by the projection of a suitable bipartite graph of individuals and
groups onto just the individuals, as shown in Fig.~\ref{model}.  The
percolation properties of the model can be solved exactly using generating
function methods.

In~\cite{Newman03e} the model was studied in a simple version in which the
size of all groups was assumed the same.  This case can account for the
presence of clustering in the network, and is straightforward to treat
mathematically.  However, it is inadequate for our purposes here, since it
does not produce any degree correlation.  Degree correlation arises because
individuals who belong to small groups tend to have low degree and are
connected to others in the same group, who also have low degree.  Similarly
those in large groups tend to have higher degree and are also connected to
one another.  Thus, the model should give rise to assortative mixing
provided there is enough variation in the sizes of groups.  As we will see,
this is indeed the case.

In addition to the parameter~$p$, we characterize the model by two
probability distributions: $r_m$~is the probability that an individual
belongs to $m$ groups and $s_n$ is the probability that a group contains
$n$ individuals.  Subject to the constraints imposed by these
distributions, the assignment of individuals to groups is entirely random.

To proceed we calculate the joint distribution~$e_{jk}$ of the excess
degrees of vertices at the ends of an edge.  Noting that the total number
of edges in groups of size~$n$ goes as $s_n n(n-1)$, we write
\begin{equation}
e_{jk} = e_0 \sum_n s_n n(n-1) P(j,k|n)
\label{ejkmodel}
\end{equation}
where $P(j,k|n)$ is the probability that an edge that belongs to a group of
size~$n$ connects vertices of excess degrees $j$ and~$k$, and $e_0$ is a
constant whose value can be calculated from the requirement that $e_{jk}$
be normalized, so that $\sum_{jk} e_{jk} = 1$.

We now decompose $j$ and $k$ in the form $j=\jin+\jout$, $k=\kin+\kout$,
where $\jin$, $\kin$ are the numbers of connections to vertices within the
group that the two vertices share, and $\jout$, $\kout$ are the numbers of
connections outside that group.  The distributions of $\jin$ and $\kin$ are
simply binomial, and hence $P(j,k|n)$ factors into terms depending only on
$j$ and $k$ thus:
\begin{eqnarray}
P(j,k|n) &=& \sum_{\jin} {n-2\choose\jin} p^{\jin} q^{n-2-\jin} P(\jout)
             \nonumber\\
         & & \hspace{-2em} {} \times
             \sum_{\kin} {n-2\choose\kin} p^{\kin} q^{n-2-\kin} P(\kout),
\label{pjkn}
\end{eqnarray}
where $P(\jout)$ is the probability distribution of $\jout$, which is
independent of~$\jin$, and similarly for $\kout$.

To evaluate this expression we introduce the following generating
functions for the distributions $r_m$ and~$s_n$:
\begin{eqnarray}
\label{defsf0f1}
\hspace{-2em} f_0(z) &=& \! \sum_{m=0}^\infty r_m z^m,\hspace{8.0pt}
              f_1(z)  =  {1\over f_0'(1)} \sum_{m=0}^\infty m r_m z^{m-1},\\
\label{defsg0g1}
\hspace{-2em} g_0(z) &=& \sum_{n=0}^\infty s_n z^n,\hspace{12.2pt}
              g_1(z)  =  {1\over g_0'(1)} \sum_{n=0}^\infty n s_n z^{n-1}.
\end{eqnarray}
Physically, $f_0(z)$~is the generating function for the number of groups an
individual belongs to, and $f_1(z)$ is the generating function for the
number groups that an individual in a randomly selected group belongs to,
other than the randomly selected group itself.  Similarly $g_0(z)$
generates the group sizes and $g_1(z)$ generates the number of other
individuals in a group to which a randomly selected individual belongs.  Of
these others, our randomly selected individual is connected to a number
binomially distributed according to the probability~$p$ and thus generated
by the simple generating function $pz+q$, where $q=1-p$.  Averaging over
the group sizes, the number of neighbors of a randomly chosen individual
within one of the groups to which they belong is generated by~$g_1(pz+q)$,
and an individual belonging to a randomly chosen group will have a number
of neighbors in other groups generated by $f_1(g_1(pz+q))$.  This then
gives us precisely the quantity $P(\jout)$ of Eq.~\eref{pjkn}, which is
equal to the coefficient of $z^{\jout}$ in $f_1(g_1(pz+q))$, and similarly
for $P(\kout)$.

Combining Eqs.~\eref{ejkmodel} and~\eref{pjkn} we find that $e_{jk}$ is
generated by the double probability generating function
\begin{eqnarray}
E(x,y) &=& \sum_{jk} e_{jk} x^j y^k \nonumber\\
       &=& g_2\bigl((px+q)(py+q)\bigr) \nonumber\\
       & & \quad {} \times f_1(g_1(px+q)) f_1(g_1(py+q)),
\end{eqnarray}
where
\begin{equation}
g_2(z) = \frac{1}{g_0''(1)} \sum_{n=0}^{\infty} n(n-1) s_n z^{n-2}.
\label{defsg2}
\end{equation}

Then, making use of Eqs.~\eref{defsqk} and~\eref{defsr} and the fact that
$q_k=\sum_j e_{jk}$, we can write the assortativity coefficient~$r$ as
\begin{equation}
r = \frac{\partial_x\partial_y E - (\partial_x E)\,(\partial_y E)}{%
    \partial_x(x\,\partial_x E)-(\partial_x E)\,(\partial_y E)}\Biggr|_{x=y=1}
    = {\mathcal{P}\over\mathcal{Q}},
\label{modelr}
\end{equation}
where the numerator and denominator $\mathcal{P}$ and $\mathcal{Q}$ are
\begin{widetext}
\begin{subequations}
\begin{eqnarray}
\mathcal{P} &=& p \mu_1^2 \nu_1^2 \bigl[(\nu_4-\nu_3)(\nu_2-\nu_1)
                -(\nu_3-\nu_2)^2\bigr] \\
\mathcal{Q} &=& \mu_1\nu_1(\nu_2-\nu_1) \bigl[(\mu_2-\mu_1)
                (\nu_2-\nu_1)^2+\mu_1\nu_1(2\nu_1-3\nu_2+\nu_3)\bigr]
                \nonumber\\
            & & \quad {} + p\bigl[(\mu_1^2-\mu_2^2-\mu_1\mu_2+\mu_1\mu_3)
                (\nu_2-\nu_1)^4+\mu_1\mu_2\nu_1 (\nu_2-\nu_1)^2
                (2\nu_1-3\nu_2+\nu_3)
                \nonumber\\
            & & \quad {} + \mu_1^2 \nu_1 \bigl(\nu_1^2
                (2\nu_2+\nu_3-\nu_4)-\nu_1(\nu_3-\nu_2)^2-\nu_1\nu_2
                (\nu_4-5\nu_2)+\nu_2^2(3\nu_2-\nu_3)\bigr)\bigr]
\end{eqnarray}
\label{pq}
\end{subequations}
\end{widetext}
In this expression the quantities $\mu_n$ and $\nu_n$ are the $n$th moments
of the distributions $r_m$ and $s_n$ respectively.  Thus, given the
distributions and the probability $p$ it is elementary, if tedious, to
calculate~$r$.  Below we apply this expression to two real-world example
networks.  First, however, a few points are worth noting.

It is straightforward to show, though certainly not obvious to the eye,
that the expression for~$r$, Eq.~\eref{modelr}, is non-negative for all
distributions $r_m$ and~$s_n$, so that our model always produces an
assortatively mixed network, as our intuition suggests.

Now consider the simple case in which each individual belongs to exactly
one group, and the group sizes have a Poisson distribution.  In this case,
Eq.~\eref{modelr} gives $r=p$, and we can achieve any value of $r$ by
tuning the parameter~$p$.  In particular, if each individual knows all
others in their group then $p=1$ and we have perfect assortativity.  This
is reasonable, since in this case each individual in a group has the exact
same number of neighbors.  This case is a rather pathological one however,
since if everyone belongs to only one group, then the network consists of
many isolated groups and most people are not connected to one another.  To
make things more realistic, let us allow the number of groups to which
individuals belong also to vary according to a Poisson distribution.  Then
we find that
\begin{equation}
r = \frac{p}{1+\mu+\nu\mu p},
\label{rpoisson}
\end{equation}
where $\mu\equiv\mu_1$ and $\nu\equiv\nu_1$ are the means of the two
distributions.  Thus as the two means increase, the correlation decreases.
The decrease with $\mu$ is easily understood---the more groups an
individual belongs to, the less the relative within-group degree
correlation upon which the assortativity depends: the within-group
correlation is diluted by all the other groups the individual belongs to.
The behavior with $\nu$ is a little more subtle.  The width of the Poisson
distribution of group sizes goes as $1/\sqrt{\nu}$ as a fraction of the
mean, and hence the effective variation in size between groups decreases
with increasing~$\nu$.  It is this decrease that drives $r$ towards zero.

\begin{figure*}
\begin{center}
\resizebox{\textwidth}{!}{\includegraphics{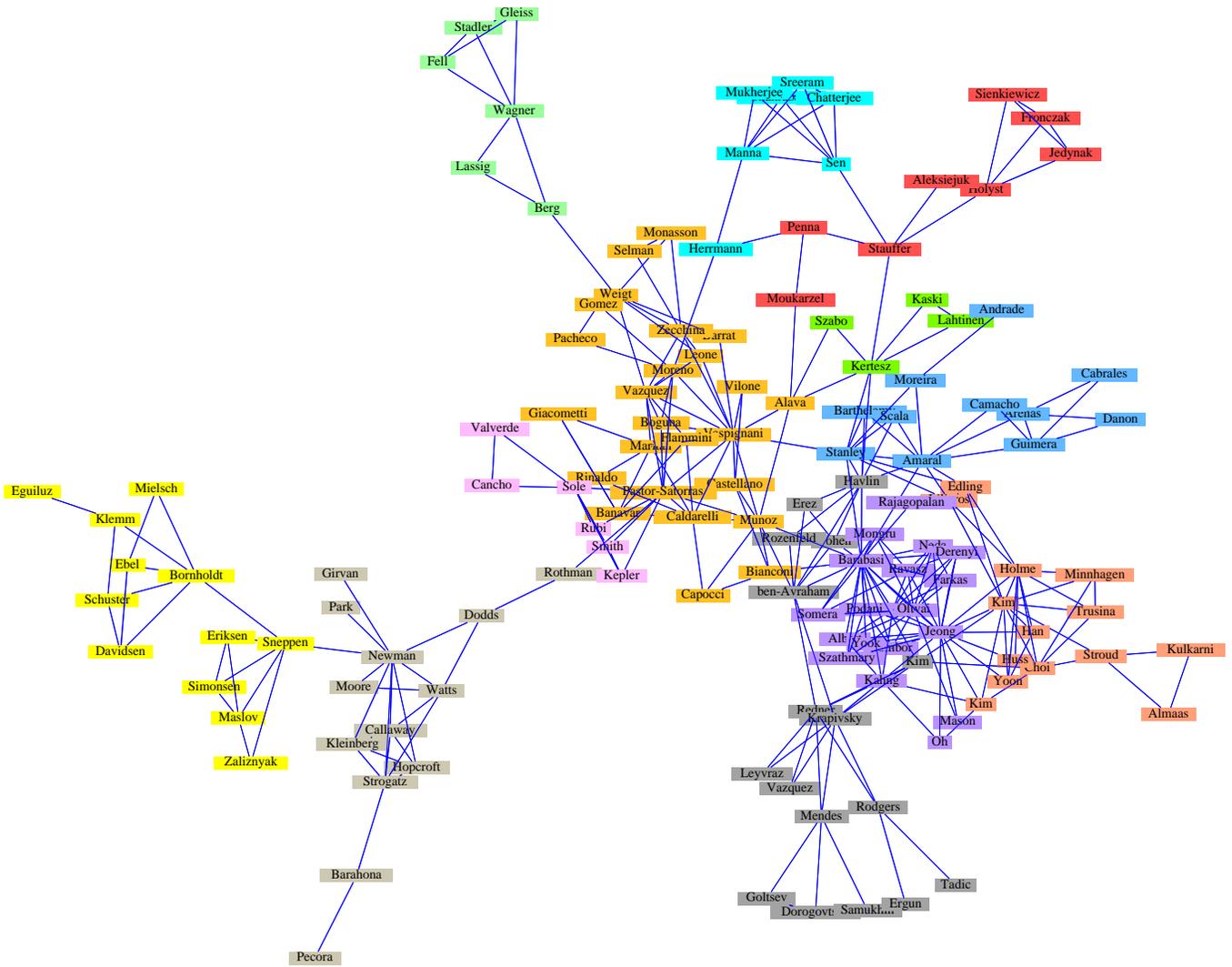}}
\end{center}
\caption{The largest component of the network of coauthorships described
in the text.  This component contains 142 scientists, and there are 36
other components, of sizes ranging from 1 to~5, containing 84 more.  The
vertices are colored according to the communities found using the algorithm
of Ref.~\cite{GN02}.  The communities correspond reasonably closely to
geographical and institutional divisions between the scientists shown.}
\label{collab}
\end{figure*}

\section{Examples}
\label{examples}
We now apply our model to two real-world example networks.  In the first
case, as we will see, it gives a value of $r$ in excellent agreement with
the real network.  In the second it underestimates $r$ by about a factor of
two, indicating that group structure can account for only a portion of the
observed assortativity, the rest, we conjecture, being due to true social
effects.

\subsection{Collaboration network}
Networks of coauthorship of scientists or other academics provide some of
the best-documented examples of social networks~\cite{GI95,Newman01a}.
Using bibliographic databases it is possible to construct large
coauthorship networks with high reliability, and these networks are true
social networks, in the sense that it seems highly likely that two authors
who write a paper together are acquainted.

Fig.~\ref{collab} shows a coauthorship network of physicists who conduct
research on networks.  The network was constructed using names drawn from
the large bibliography of Ref.~\cite{Newman03d} and coauthorship data from
preprints submitted to the condensed matter section of the Physics E-print
Archive at \verb|arxiv.org| between Jan~1, 1995 and April~30, 2003.  To
find the groups in the network, we fed it through the community structure
algorithm of Girvan and Newman~\cite{GN02}, producing the division shown by
the colors in the figure.  The figure shows only the largest component of
the network.  There are also 36 smaller components, which were included in
our calculations even though they are not shown.

The moments of the distributions $r_m$ and $s_n$ are easily extracted from
the network by direct summation.  To find the value of~$p$, we counted the
number of edges in the network and divided by the total number of possible
within-group edges, giving $p=0.178$.  Feeding this value and our figures
for the moments into Eqs.~\eref{modelr} and~\eref{pq}, we then find a
predicted value of $r=0.145$.  The measured value for the real network is
$0.174\pm0.045$.  (The error is calculated according to the prescription
given in~\cite{Newman03c}.)  These two figures are in agreement within the
statistical error on the latter.

While this result by no means proves that the group structure is
responsible for assortativity in this network, it tells us that no other
assumption is necessary to give the observed value of~$r$.  With group
structure as shown in the figure and otherwise random mixing, we would get
a network with exactly the assortativity that is observed in reality,
within expected error.

\subsection{Boards of directors}
Davis and collaborators~\cite{DG97,DYB01} have studied networks of the
directors of companies in which two directors are considered connected if
they sit on the board of the same company.  They studied the Fortune 1000,
the one thousand US companies with the highest revenues, for 1999, and
assembled a near-complete director network from publicly available data.
The network consists of 7673 directors sitting on 914 boards.  It provides
a particularly simple example of our method, for two reasons.  First, the
groups in the network through which individuals are acquainted are provided
for us---they are the boards of directors.  Second, it is assumed that
directors are acquainted with all those with whom they share a board, so
that the parameter~$p$ in our model is~1.

The distributions of boards per director and directors per board have been
studied before~\cite{NSW01}.  We note that most directors (79\%) sit on
only one board and that there is considerable variation in the size of
boards (from 2 to 35 members).  Thus we would expect strong assortative
mixing in the network, and indeed we find that $r=0.276\pm0.004$.  Taking
the moments of the measured distributions $r_m$ and $s_n$ for the network
and setting $p=1$, Eq.~\eref{modelr} gives a value of $r=0.116$ for our
model.  So it appears that the presence of groups in the network can
explain about 40\% of the assortativity we observe in this case, but not
all of it.  There is some additional assortativity in addition to the
purely topological effect of the groups, and we conjecture that this is due
to some true sociological or psychological effect in the way in which
acquaintanceships are formed.  One possibility is suggested by the analysis
of the directorships data by Newman~\etal~\cite{NSW01}, who found that
directors who sit on many boards tend to sit on them with others who sit on
many boards.  Since those who sit on many boards will also tend to have
high degree, we would expect this effect to add assortativity to the
network, but the effect is missing from our model in which board membership
is assigned at random.

In a sense, our model is giving a baseline against which to measure the
value of~$r$; it tells us when the value we see is simply what would be
expect by random chance, as in the collaboration network above, and when
there must be additional effects at work, as in the boards of directors.

\section{Conclusions}
\label{concs}
In this paper we have argued that social and non-social networks differ in
two important ways.  First, they show distinctly different patterns of
correlation between the degrees of adjacent vertices, with degrees being
positively correlated (assortative mixing) in most social networks and
negatively correlated (disassortative mixing) in most non-social networks.
Second, social networks show high levels of clustering or network
transitivity, whereas clustering in many non-social networks is no higher
than one would expect on the basis of pure chance, given the observed
degree distribution.

We have shown that both of these differences can be explained by the same
hypothesis, that social networks are divided into communities, and
non-social networks are not.  We have studied a simple model of community
structure in social networks in which individuals belong to groups and are
acquainted with others with whom they share those groups.  The model is
exactly solvable using generating function techniques, and we have shown
that it gives predictions that are in reasonable and sometimes excellent
agreement with empirical observations of real-world social networks.

\begin{acknowledgments}
The authors would like to thank Duncan Watts for useful conversations.
This work was funded in part by the National Science Foundation under grant
number DMS--0234188 and by the James S. McDonnell Foundation and the Santa
Fe Institute.
\end{acknowledgments}

\end{document}